\documentclass[aps,prl,reprint,preprintnumbers,superscriptaddress,amsmath,amssymb,bibnotes,longbibliography]{revtex4-1}

\usepackage{graphicx}
\usepackage{dcolumn}
\usepackage{bm}
\usepackage[colorlinks,linkcolor=blue,anchorcolor=blue,citecolor=blue,urlcolor=blue,filecolor=blue,menucolor=blue,runcolor=blue]{hyperref}
\usepackage{ulem}

\begin{document}

\title{A single full gap with mixed type-I and type-II superconductivity on surface of the type-II Dirac semimetal PdTe$_2$ by point-contact spectroscopy}

\author{Tian Le}
\author{Lichang Yin}
\affiliation  {Center for Correlated Matter and Department of Physics, Zhejiang University, Hangzhou 310058, China}
\author{Zili Feng}
\affiliation  {Beijing National Laboratory for Condensed Matter Physics, Institute of Physics, Chinese Academy of Sciences, Beijing 100190, China}
\author{Qi Huang}
\author{Liqiang Che}
\author{Jie Li}
\affiliation  {Center for Correlated Matter and Department of Physics, Zhejiang University, Hangzhou 310058, China}

\author{Youguo Shi}
\affiliation  {Beijing National Laboratory for Condensed Matter Physics, Institute of Physics, Chinese Academy of Sciences, Beijing 100190, China}
\affiliation  {School of Physical Sciences, University of Chinese Academy of Sciences, Beijing 100049, China}
\author{Xin Lu}
\email[Corresponding author: ]{xinluphy@zju.edu.cn}
\affiliation  {Center for Correlated Matter and Department of Physics, Zhejiang University, Hangzhou 310058, China}
\affiliation  {Zhejiang Province Key Laboratory of Quantum Technology and Device, Zhejiang University, Hangzhou 310027, China}
\affiliation{Collaborative Innovation Center of Advanced Microstructures, Nanjing University, Nanjing, 210093, China}

\date{\today}

\begin{abstract}
We report our point-contact spectroscopy (PCS) study on the superconducting state of the type-II Dirac semimetal PdTe$_2$ with a superconducting transition temperature $T\rm_c \sim$ 1.65 K. Both mechanical- and soft- PCS differential conductance curves at 0.3 K show a consistent double-peak structure and they can be perfectly fitted by a single s-wave gap based on the Blonder-Tinkham-Klapwijk model. The gap follows a typical Bardeen-Cooper-Schrieffer temperature behavior, yielding $\Delta_0 \sim$ 0.29 meV and 2$\Delta_0$/$k\rm_{B}$$T\rm_{c}$ = 4.15 in the strong coupling regime. A sudden suppression of the superconducting gap in magnetic field around $H{\rm_{c1}}$ $\sim$ 130 Oe is observed for most point-contacts on PdTe$_2$, characteristic of a first-order transition for type-I superconductor in field. However, for other contacts, a smooth evolution of the PCS conductance persists up to $H{\rm_{c2}}$ $\sim$ 600 Oe, signaling a local type-II superconductivity. The observed admixture of type-I and type-II superconductivity can possibly arise from an inhomogeneous electron mean free path on the surface of PdTe$_2$ due to its topological surface states.
\end{abstract}

\maketitle

Soon after the discovery of topological insulators (TIs), topological superconductors (TSCs) have also attracted intensive attention in the community because of the nontrivial topology for Bogoliubov  quasiparticles in TSCs and electronic bands in TIs \cite{RevModPhys.83.1057, RevModPhys.82.3045, 031214-014501, 0034-4885-80-7-076501, PhysRevB.61.10267, PhysRevLett.102.187001, PhysRevB.78.195125, doi:10.1063/1.3149495}. Moreover, Majorana fermions are expected to be hosted on the surface or edge of TSCs, complying with non-Abelian statistics and playing a decisive role in quantum computation \cite{RevModPhys.80.1083, KITAEV20032, 030212-184337}. One strategy to realize TSCs is to induce superconductivity (SC) in topological materials, such as topological insulator, Weyl and Dirac semimetals, by taking advantage of the nontrivial topology of the electronic bands \cite{PhysRevB.79.214526, PhysRevLett.105.097001, PhysRevB.81.134508, PhysRevB.85.180509, PhysRevLett.114.096804, PhysRevLett.115.187001, PhysRevB.92.035153}. For example, topological SC has been claimed for the topological surface states with proximity-induced SC at the interface of a TI/SC heterostructure or on the surface of some iron-based superconductors, where Majorana zero modes are argued to exist at their vortices or edges \cite{PhysRevLett.100.096407, PhysRevLett.114.017001, xu2014momentum, Wang52, PhysRevLett.112.217001, PhysRevLett.112.217001, zareapour2012proximity, wang2013fully, Zhang182, Wangeaao1797, PhysRevX.8.041056}.

Recently, the transition-metal dichalcogenide compound PdTe$_2$ has been confirmed to be a type-II Dirac semimetal with a tilted Dirac cone below the Fermi energy by the angle resolved photoemission spectroscopy (ARPES) and there exists a spin-polarized topological surface state \cite{PhysRevLett.119.016401, PhysRevLett.120.156401}. De Haas-van Alphen (dHvA) experiments also evidence the nontrivial Berry phase for one of the Fermi-surface pockets, probably a hole pocket from the tilted Dirac cone \cite{PhysRevB.96.041201, PhysRevB.97.235154}. On the other hand, superconductivity in PdTe$_2$ with a transition temperature $T\rm_{c}$ $\sim$ 1.7 K has been known for a long time, serving as a promising candidate for stoichiometric TSCs \cite{PhysRevB.96.220506}. However, heat capacity, London penetration depth, tunneling junction and STM measurements all support a conventional fully-gapped s-wave superconductor for PdTe$_2$ \cite{PhysRevB.97.054515, PhysRevB.98.024508, Salis_2018, Sirohi_2019, PhysRevLett.120.156401, PhysRevB.97.014523}. Moreover, recent electrical transport and heat capacity measurements reveal a puzzling discrepancy of its critical field for the superconducting state with 3000 Oe and 250 Oe, respectively. It is difficult to be explained by either a general Saint-James-de Gennes (SJdG) surface critical field or filamentary superconductivity \cite{PhysRevB.96.220506, PhysRevB.97.054515}. One possible scenario proposed by Siroshi \textit{et al.} is that bulk PdTe$_2$ is a typical type-I superconductor and there exists a mixing of type-I and type-II superconductivity on its surface \cite{PhysRevB.96.220506, PhysRevB.97.054515, Sirohi_2019}.  More experiments are thus desired to elucidate its exact superconducting nature for the type-II Dirac semimetal PdTe$_2$.

In this article, we have applied both mechanical and soft point-contact spectroscopy (PCS) to investigate the superconducting gap of single crystalline PdTe$_2$. Our PCS conductance curves at 0.3 K can be well fitted by a single gap s-wave Blonder-Tinkham-Klapwijk (BTK) model \cite{PhysRevB.25.4515} and the temperature evolution of the extracted superconducting gap follows a conventional Bardeen-Cooper-Schrieffer (BCS) behavior, yielding a gap $\Delta_0 \sim$ 0.29 meV with 2$\Delta_0$/$k\rm_{B}$$T\rm_{c}$ = 4.15. Even though some point-contacts show an abrupt suppression of the SC gap by magnetic field at 130 Oe, characteristic of a type-I superconductor, some other contacts exhibit a smooth evolution in field with a typical type-II superconducting behavior below a critical field of 600 Oe, suggesting an inhomogeneous mixing of type-I and type-II SC on the PdTe$_2$ crystal surface.

\begin{figure}
\includegraphics[angle=0,width=0.49\textwidth]{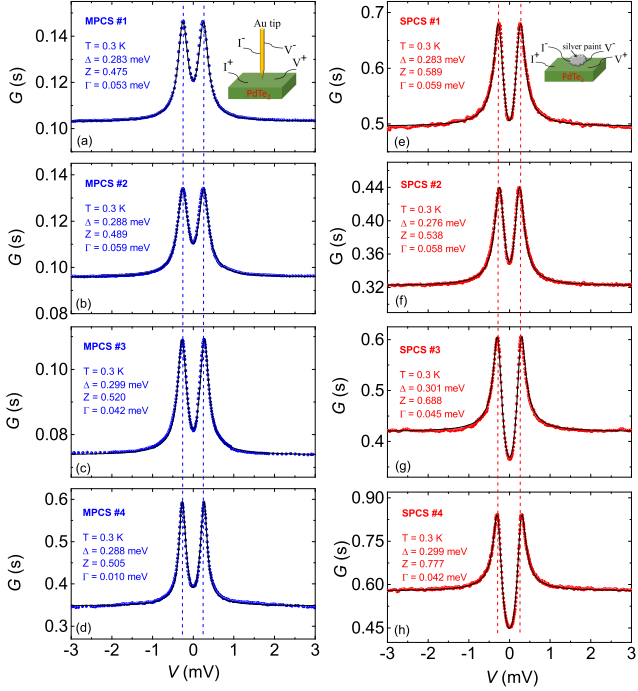}
\vspace{-12pt} \caption{\label{Figure1}(color online) A representative set of  point-contact conductance curves on PdTe$_2$ at 0.3 K for different contacts from MPCS (a-d) and SPCS (e-h) in comparison with their optimized single gap s-wave BTK fittings (black lines). The insets in (a) and (e) are schematic illustrations for MPCS and SPCS, respectively.}
\vspace{-12pt}
\end{figure}

PdTe$_2$ single crystals were grown by the flux method: High purity Pd and Te with the molar ratio 1:4 were put into an aluminum crucible and then sealed in a quartz tube. The sample was heated up to 800 $^\circ$C and then cooled to 500 $^\circ$C slowly. The excess Te were removed by centrifuge at this temperature and single crystals PdTe$_2$ were left at the bottom of the crucible. Before experiment, the samples were cleaved to expose fresh surface at ambient condition. Mechanical PCS (MPCS) in a needle-anvil configuration was employed, where an electrochemically-etched sharp gold tip was gently engaged on the crystal surface by piezo-controlled nano-positioners. In comparison, soft PCS (SPCS) contacts on PdTe$_2$ were prepared on the sample surface by attaching a 30 $\mu$m diam gold wire with a silver-paint drop at the end yielding a total contact diameter around $\sim$ 50-100 $\mu$m. Hundreds of parallel nanoscale junctions are assumed between individual silver particles and the crystal surface for SPCS \cite{Daghero_2010, PhysRevLett.107.217001}. In the case of MPCS, even though parallel nanoscale channels should also exist, the total contact diameter from the sharp tip end is less than 5 $\mu$m, much smaller than that of SPCS. The PCS differential conductance curves as a function of bias voltage, $G$($V$), were recorded with the conventional lock-in technique in a quasi-four-probe configuration. Oxford cryostat with a He3 insert was used to cool the whole PCS setup down to 0.3 K and magnetic field was applied along the crystal c-axis up to 1000 Oe to suppress SC.

\begin{figure}
\includegraphics[angle=0,width=0.49\textwidth]{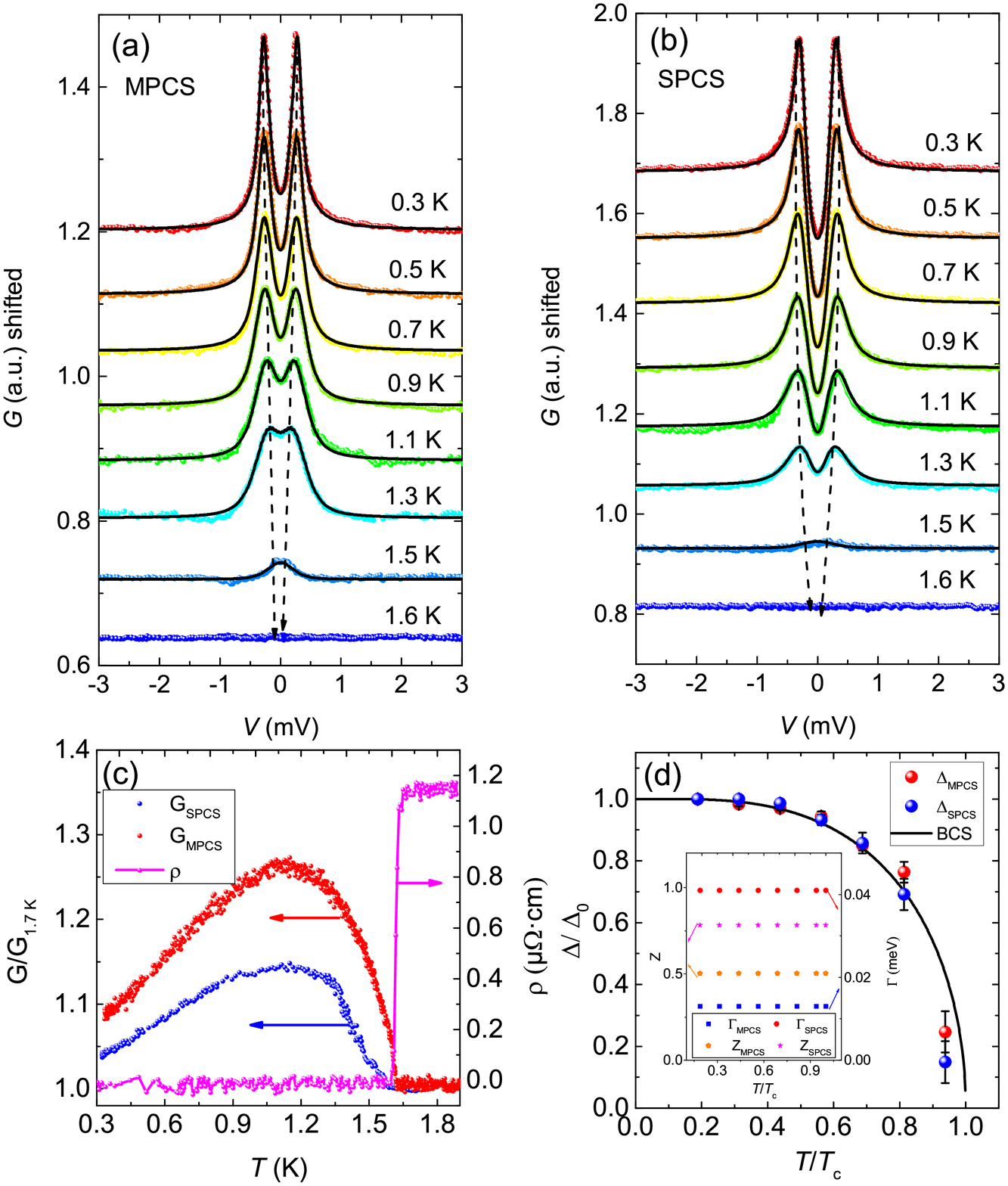}
\vspace{-12pt} \caption{\label{Figure2}(color online) (a) and (b) Temperature evolution of the differential conductance curves $G$($V$) from 0.3 K to 1.6 K for MPCS and SPCS on PdTe$_2$, respectively, in comparison with a single gap s-wave BTK fitting (black lines). (c) Temperature dependence of the electrical resistivity $\rho$ (pink dots) and normalized zero-bias conductance $G/G_{1.7 K}$ curves from MPCS (red dots) and SPCS (blue dots), respectively. (d) Temperature dependence of the extracted superconducting gap $\Delta$ from the single-gap BTK fitting for both MPCS and SPCS in accordance with the BCS curve. The inset shows little change of fitting parameters Z and $\Gamma$ in MPCS and SPCS as a function of temperature.}
\vspace{-12pt}
\end{figure}

Figure \ref{Figure1} shows eight representative PCS differential conductance curves $G$($V$) on PdTe$_2$ at the lowest temperature 0.3 K with the left and right panel for MPCS and SPCS, respectively. All the  conductance curves show a common double-peak structure around $\pm$ 0.3 meV and no dip feature at high bias is present in any curve, ensuring the ballistic nature of our contacts \cite{PhysRevB.69.134507}. A single gap s-wave BTK model can perfectly fit all experimental curves as shown by the black lines in Fig. \ref{Figure1} and the extracted superconducting gap $\Delta$ values at 0.3 K are scattered in the range of 0.276 and 0.301 meV, yielding 2$\Delta$/$k\rm_{B}$$T\rm_{c}$ = 4.00 - 4.36 in a strong-coupling regime. We notice that our PCS gap values are consistent with those reported by STM ($\sim$ 0.289 meV) \cite{Sirohi_2019, PhysRevB.97.014523}, but larger than the weak-coupling value estimated from bulk specific heat and penetration depth studies \cite{PhysRevB.97.054515, PhysRevB.98.024508}. The discrepancy of gap size is suspected to be associated with the spin-polarized topological surface state, because both PCS and STM are more sensitive to the SC on surface rather than the bulk SC. However, its exact origin needs further clarification \cite{LV2017852, PhysRevB.90.184516, guan2019experimental}. For optimal fittings, the smearing parameter $\Gamma$ is much smaller than the superconducting gap and comparable between MPCS and SPCS, while the barrier strength parameter Z ranges from $\sim$ 0.5 to 0.78 for different contacts in Fig. \ref{Figure1} possibly from the Fermi velocity mismatch between the tip and sample \cite{PhysRevB.25.4515, PhysRevB.27.112}.

\begin{figure}
\includegraphics[angle=0,width=0.49\textwidth]{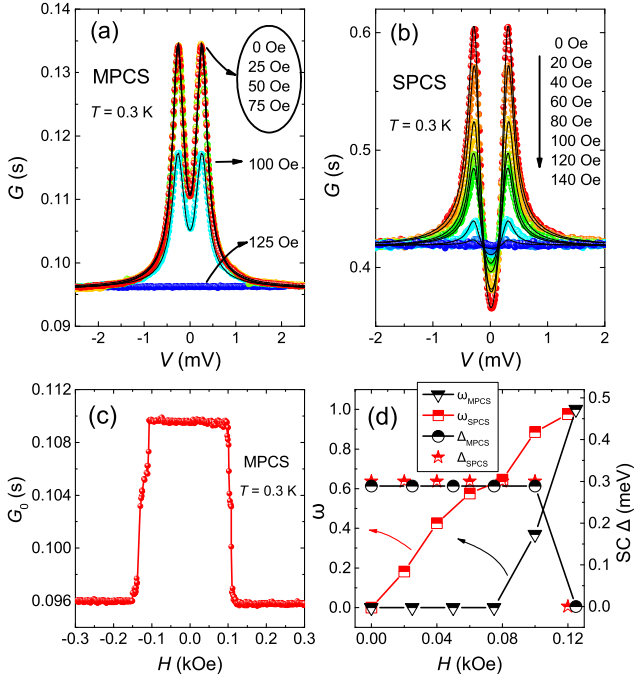}
\vspace{-12pt} \caption{\label{Figure3}(color online) (a) and (b) Point-contact differential conductance curves $G$($V$) at 0.3 K as a function of magnetic field for MPCS and SPCS on PdTe$_2$, respectively, in comparison with a two-component BTK fitting (black lines). (c) Field dependence of the MPCS zero-bias conductance has a clear first-order transition and the residual magnetic field is estimated 15 Oe. (d) Field evolution of the extracted spectra weight $\omega$ for normal state and the superconducting gap $\Delta$ for SC regions for both MPCS and SPCS.}
\vspace{-12pt}
\end{figure}

Figure \ref{Figure2}(a) and (b) show the temperature evolution of conductance curves $G$($V$) from 0.3 to 1.6 K for MPCS and SPCS on PdTe$_2$, respectively. With increased temperatures, the double peaks are smeared to a single peak and finally disappear at the superconducting  $T\rm_c \sim$ 1.65 K for PdTe$_2$. We note that the $T\rm_c$ determined by the zero-bias conductance (ZBC) curves for both mechanical and soft contacts are around 1.65 K, consistent with the resistive $T\rm_c$ as shown in Fig. \ref{Figure2}(c). The extracted SC gap values from an optimal BTK fitting are shown in Fig. \ref{Figure2}(d) and it follows the conventional BCS temperature behavior. As in the inset of Fig. \ref{Figure2}(d), the fitting parameters Z and $\Gamma$ for both MPCS and SPCS show little change with temperature, indicating an ideal ballistic contact \cite{Huang_2008}.

\begin{figure}
\includegraphics[angle=0,width=0.49\textwidth]{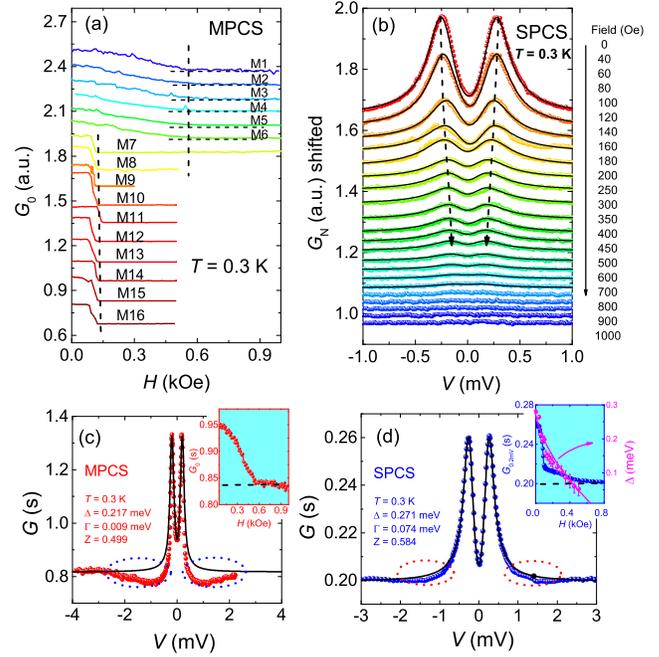}
\vspace{-12pt} \caption{\label{Figure4}(color online) (a) Field dependence of the zero-bias conductance for different MPCS contacts on the surface of PdTe$_2$. (b) One set of SPCS conductance curves as a function of field show a type-II SC behavior at 0.3 K in comparison with the standard BTK fitting (black lines). (c)-(d) The typical conductance curve $G$($V$) for a type-II SC behavior at 0.3 K has a dip structure at high bias for MPCS and SPCS, respectively. The insets show the field-dependent contact conductance at zero-bias for MPCS and fixed bias voltage 0.2 meV for SPCS, respectively, while the extracted SC gap from the standard BTK fitting has a smooth evolution with field.}
\vspace{-12pt}
\end{figure}

For the MPCS field dependence, the conductance curves $G$($V$) show an interesting behavior as in Fig. \ref{Figure3}(a): They barely change but overlap with each other below 75 Oe, and the conductance peaks are abruptly suppressed in intensity at 100 Oe and quickly disappear around 125 Oe, while the peak positions stay almost unshifted in field. The MPCS zero-bias conductance as a function of field displays a sudden drop at 130 Oe as in Fig. \ref{Figure3}(c), strongly supporting a type-I superconductor for PdTe$_2$ as reported by magnetization, heat capacity and STM measurements \cite{PhysRevB.96.220506, PhysRevB.97.054515, Sirohi_2019}. In contrast, the SPCS conductance curves for PdTe$_2$ as in Fig. \ref{Figure3}(b) show a  gradually reduced peak intensity in field, whereas the peak positions have no change before the suppression of SC at 130 Oe. We note that, for type-I superconductors, an intermediate state will emerge far below its critical field when the field is perpendicular to the sample plane with a large demagnetization factor. In such a case, normal state domains can coexist with the SC domains in space and progressively replace the superconducting volume \cite{PhysRevLett.98.257001, PhysRevB.89.100503, PhysRevB.72.212508}. In order to mimic this process, a modified two-component BTK model has been applied to fit the conductance curves with both contributions from the normal and SC regions, $G(V)$ = $\omega$$G_{Normal}$ + (1-$\omega$)$G\rm_{BTK}$$(V)$, where $G_{Normal}$ is just flat for the normal state, $G\rm_{BTK}$$(V)$ is the standard BTK curve for the SC regions, and the fitting parameter $\omega$ denotes the spectra weight from the normal state. The SPCS conductance curves at different fields can be well fitted by this modified BTK model as shown in Fig. \ref{Figure3}(b) and the parameters Z and $\Gamma$ change little with field as in MPCS. The normal state spectra weight $\omega$ shows a roughly linear increase with field as in Fig. \ref{Figure3}(d) and vividly illustrates a monotonic increase of the normal state volume in the intermediate state. Nonetheless, the SC gap for the SC regions keeps constant in field up to 100 Oe for PdTe$_2$, indicative of a type-I superconductor. This modified BTK model can also be applicable to the MPCS case as in Fig. \ref{Figure3}(d): A sudden increase of the $\omega$ value from zero implies normal state domains entering the contact area, only when the field gets close to $H{\rm_{c1}}$. The systematic difference between MPCS and SPCS is due to its much smaller total contact area in MPCS so that the tip only probes a single superconducting domain for MPCS in low fields.

For MPCS, we have collected a set of ZBC curves as a function of field at 0.3 K, $G_0(H)$, for contacts M1-M16 as shown in Fig. \ref{Figure4}(a). Each contact is obtained once after the tip is withdrawn, moved to another position by xy positioners and gently engaged on the sample with the z positioner, where M1-M16 are from the same tip but not numerically ordered as the experimental sequence. While type-I SC for PdTe$_2$ can been established by the sudden drop at 130 Oe in ZBC for contacts M7-M16, we notice that other ZBC curves M1-M6 show a smooth evolution into the normal state at a critical field $H\rm_{c2}$ around 600 Oe. The continuous transition for M1-M6 with a much larger critical field than the bulk $H{\rm_{c1}}$ $\sim$ 130 Oe suggests a local type-II superconductivity at these contacts, consistent with the STM and resistivity measurements for current along c axis \cite{Sirohi_2019, PhysRevB.96.041201}. A corresponding conductance curve $G(V)$ at 0.3 K has an obvious dip structure at higher bias voltages as in Fig. \ref{Figure4}(c), implying the contact not in a pure ballistic limit and the mean free path $l$ is believed smaller than the diameter of each contact channel \cite{PhysRevB.69.134507}. It is intriguing that no additional dips are observed for any contact among M7- M16 with type-I SC, while dips are commonly observed for contacts M1-M6 with type-II SC. Some SPCS conductance curves also exhibit a smooth evolution of the SC gap with field as in Fig. \ref{Figure4}(b) and the inset of Fig. \ref{Figure4}(d). Its conductance at a fixed bias voltage 0.2 meV shows a similar continuous type-II behavior at 600 Oe after a first-order-like drop at 130 Oe as in the inset of Fig. \ref{Figure4}(d), signaling a spacial coexistence of type-I and type-II SC in the soft contact area (We note that the zero-bias conductance for our SPCS contacts available is comparable to the value in the normal state, and so the conductance at a fixed bias voltage 0.2 meV rather rather than zero bias is chosen to easily identify the critical field). A weak dip structure at high bias is also observed with the conductance curve deviating from its BTK fitting as in Fig. \ref{Figure4}(d). We notice that a similar contact-dependence for PCS spectra has been reported in a recent study on the non-centrosymmetric superconductor Re$_6$Zr and it is ascribed to the surface superconductivity as well \cite{parab2019point}.

\begin{figure}
\includegraphics[angle=0,width=0.49\textwidth]{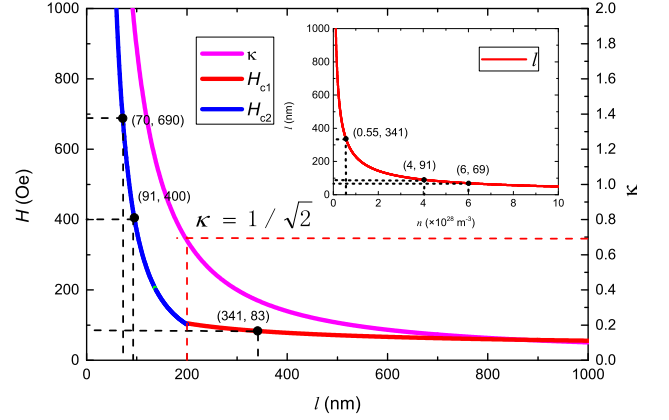}
\vspace{-12pt} \caption{\label{Figure5}(color online) Mean free path $l$ dependence of the Ginzburg-Landau parameter $\kappa$ (pink curve) and the corresponding critical fields for type-I and type-II superconductor regimes (red and blue segments, respectively). The inset shows the calculated mean free path $l$ as a function of carry density $n$ based on the Drude model.}
\vspace{-12pt}
\end{figure}

In general, the Ginzburg-Landau (GL) parameter $\kappa$ = $\frac{\lambda}{\xi}$ is defined to describe type-I ($\kappa$ $<$ $\frac{1}{\sqrt{2}}$ ) and type-II ($\kappa$ $>$ $\frac{1}{\sqrt{2}}$) superconductors, where $\lambda$ is the penetration depth and $\xi$ the coherence length of superconductors \cite{tinkham2004introduction}. It was claimed that the $\kappa$ in PdTe$_2$ is around 0.34 $<$ $\frac{1}{\sqrt{2}}$, categorizing PdTe$_2$ as a type-I superconductor \cite{PhysRevB.96.220506}. In order to explain the emergence of type-II SC behavior, we follow the arguments by Siroshi \textit{et al.} and consider the Pippard nonlocal electrodynamics $\frac{1}{\xi} = \frac{1}{\xi_0} + \frac{1}{l}$ and $\lambda = \lambda_L\sqrt{1+\frac{\xi_0}{l}}$, where $\xi_{0}$ and $\lambda_L$ is the BCS coherence length and London penetration depth with $l$ $\rightarrow\infty$, respectively. Following the relations $\xi_0 = \frac{0.18\hbar^2k_F}{k_BT_cm^\ast}$ and $\lambda_L = \sqrt{\frac{m^\ast}{\mu_0ne^2}}$, if we assume the carrier density n = 5.5 $\times 10^{27}$ m$^{-1}$\cite{PhysRevB.96.220506}, the effective mass $m^\ast$ $\approx0.3m_e$\cite{Dunsworth1975, PhysRevB.96.041201} ($m_e$ is the free electron mass), and the Fermi wave number $k_F = \sqrt[3]{3\pi^2n}$ = 5.5$\times 10^9$ m$^{-1}$, we can get $\lambda_L = 39$ nm and $\xi_0 = 1800$ nm for PdTe$_2$. Since $\kappa = \lambda_L(\frac{1}{\xi_0}+\frac{1}{l})\sqrt{1+\frac{\xi_0}{l}}$, $\kappa$ as a function of $l$ is plotted in Fig. \ref{Figure5} as the pink line. We note that the mean free path $l$ for bulk PdTe$_2$ is estimated $l\approx$ 341 nm from the Drude model $l = \frac{3\pi^2\hbar}{\rho e^2k{_F^2}}$, which makes $\kappa \sim$ 0.341 in the type-I superconductor regime. A reduced mean free path $l$ on some surface regions can cause an enhanced $\kappa$ and thus tune it into a type-II SC with $l\leq$ 200 nm. We thus argue that the dip structure at high bias frequently observed for contacts with the type-II SC behavior implies a greatly reduced mean free path $l$, which makes local surface SC regions enter into the type-II regime.

As for the critical field, according to the GL theory, $H{\rm_{c1}} = \frac{\phi_0}{2\sqrt{2}\pi\mu_0\lambda\xi} = \frac{\phi_0}{2\sqrt{2}\pi\mu_0\lambda_L\sqrt{\xi_0}}\sqrt{\frac{1}{\xi_0}+\frac{1}{l}}$ and $H{\rm_{c2}} = \frac{\phi_0}{2\pi\mu_0\xi^2} = \frac{\phi_0}{2\pi\mu_0}(\frac{1}{\xi_0}+\frac{1}{l})^2$. The dependence of $H{\rm_{c1}}$ and $H{\rm_{c2}}$ on the mean free path $l$ is shown in Fig. \ref{Figure5} with red and blue segments, respectively. A mean free path $l$ $\sim$ 341 nm estimated from the Drude model will give an $H{\rm_{c1}}$ around 83 Oe not so off from our PCS value 130 Oe for type-I SC behavior. As shown in Fig. \ref{Figure5}, a slight variation of the mean free path $l$ has no dramatic effect on $H{\rm_{c1}}$ in the type-I regions, consistent with the fixed critical fields $\sim$ 130 Oe   from different PCS contacts.

For the type-II superconductor behavior among our PCS contacts, a critical field around 600 Oe is consistently observed without a large variation either. However, Fig. \ref{Figure5} indicates a slight change of the mean free path $l$ on surface should cause a wide distribution of $H{\rm_{c2}}$ for $l \leq$ 200 nm. Our observations probably argues against the mixture of type-I and type-II SC simply due to the disorder effect on surface. We notice that a variation of local density of states is frequently observed on the surface of several topological materials such as Bi$_2$Se$_3$, Bi$_2$Te$_3$ and BiSbTeSe$_2$, where charge puddles are proposed from a local fluctuation of the chemical potential when the Dirac cone is close to it \cite{PhysRevB.96.195135, PhysRevB.93.245149, zhang2009origin, beidenkopf2011spatial}. The accumulated charge puddles on the surface should increase the local charge carriers substantially for the low-carrier-density semimetal and thus reduce the mean free path $l$ with $l = \frac{3\pi^2\hbar}{\rho e^2k{_F^2}}$ and $k_F = \sqrt[3]{3\pi^2n}$. Once the carrier density $n$ goes beyond 4000 $\times 10^{25}$ m$^{-3}$, the reduced mean free path $l$ $\sim$ 100 nm inside different puddles doesn't change too much as shown in the inset of Fig. \ref{Figure5}, giving a critical field $H{\rm_{c2}}$ between 400 -700 Oe. This crude model can roughly explain the disparate $H{\rm_{c}}$ values between type-I and type-II SC in our PCS results.

Our MPCS and SPCS spectra show a reproducible double peak structure with the absence of a zero-bias conductance peak and a perfect fitting with a single gap s-wave BTK model strongly supports a conventional s-wave paring gap in the type-II Dirac semimetal PdTe$_2$ as evidenced by other measurements \cite{PhysRevB.97.054515, PhysRevB.98.024508, Salis_2018, Sirohi_2019, PhysRevLett.120.156401, PhysRevB.97.014523}. On the other hand, SC with a full gap is claimed for topological surface states to host Majorana zero modes in vortices for some iron-based superconductors \cite{Zhang182, Wangeaao1797, PhysRevX.8.041056}. As a type-I superconductor, it is thus difficult to check Majorana zero modes for PdTe$_2$ without vortex lattice. However, a mixture of type-I and type-II SC on the surface offers a chance to search for Majorana fermions in PdTe$_2$, if any. For PdTe$_2$, this mixing is attributed to an inhomogeneous electron mean free path $l$ on the surface. More studies are needed to illuminate the relationship between this puzzling inhomogeneous behavior and topological surface states in PdTe$_2$ especially by STM measurements.

In conclusion, we have observed a single full superconducting gap $\Delta$ from both MPCS and SPCS on the type II Dirac semimetal PdTe$_2$ with $\Delta_0 \sim$ 0.29 meV and 2$\Delta_0$/$k\rm_{B}$$T\rm_{c}$ = 4.15 in the strong coupling regime. However, the field-dependent conductance curves suggest a mixture of type-I and type-II SC with distinct critical fields probably due to an inhomogeneous electron mean free path on the surface of PdTe$_2$. More careful studies are called for to address its origin and possible relations with topological surface states in PdTe$_2$.

We are grateful for valuable discussions with L. Li, Y. Zheng and Y. Zhou. Our work was supported by the National Key Research \& Development Program of China (Grant No. 2016FYA0300402, 2017YFA0302901 and 2017YFA0303101) and the National Natural Science Foundation of China (Grant No. 11674279, 11774399 and 11374257). X.L. would like to acknowledge support from the Zhejiang Provincial Natural Science Foundation of China (LR18A04001) and Y.G.S. acknowledges support from Beijing Natural Science Foundation (Z180008).

%

\end{document}